\documentclass[doublecol,amsmath,amssymb]{epl2}
\usepackage{amssymb}
\usepackage{graphicx}
\usepackage{dcolumn}
\usepackage{bm}
\usepackage{hhline}
\usepackage{longtable}
\usepackage[usenames]{color}

\begin{document}

\title{Anomalous behavior of trapping on a fractal scale-free network}
\shorttitle{Anomalous behavior of trapping on a fractal scale-free
network}

\author{Zhongzhi Zhang\inst{1,2} \footnote{ \email{zhangzz@fudan.edu.cn} } \and Wenlei Xie\inst{1,2}  \and  Shuigeng Zhou\inst{1,2} \footnote{\email{sgzhou@fudan.edu.cn}}  \and Shuyang Gao \inst{1,2} \and Jihong Guan\inst{3} \footnote{\email{jhguan@tongji.edu.cn}}}
\shortauthor{Zhongzhi Zhang, Wenlei Xie, Shuigeng Zhou, Shuyang Gao,
and Jihong Guan}

 \institute{
  \inst{1} School of Computer Science, Fudan University, Shanghai 200433, China\\
  \inst{2} Shanghai Key Lab of Intelligent Information Processing, Fudan University, Shanghai 200433, China\\
 \inst{3} Department of Computer Science and Technology,
Tongji University, 4800 Cao'an Road, Shanghai 201804, China}

\date{\today}

\begin{abstract}{
It is known that the heterogeneity of scale-free networks helps
enhancing the efficiency of trapping processes performed on them. In
this paper, we show that transport efficiency is much lower in a
fractal scale-free network than in non-fractal networks. To this
end, we examine a simple random walk with a fixed trap at a given
position on a fractal scale-free network. We calculate analytically
the mean first-passage time (MFPT) as a measure of the efficiency
for the trapping process, and obtain a closed-form expression for
MFPT, which agrees with direct numerical calculations. We find that,
in the  limit of  a large network order $V$, the MFPT $\langle T
\rangle$ behaves superlinearly as $\langle T \rangle \sim
V^{\frac{3}{2}}$ with an exponent $\frac{3}{2}$ much larger than 1,
which is in sharp contrast to the scaling $\langle T \rangle \sim
V^{\theta}$ with $\theta \leq 1$, previously obtained for
non-fractal scale-free networks. Our results indicate that the
degree distribution of scale-free networks is not sufficient to
characterize trapping processes taking place on them. Since various
real-world networks are simultaneously scale-free and fractal, our
results may shed light on the understanding of trapping processes
running on real-life systems. }
\end{abstract}



 \maketitle

\section{Introduction}

In the past decade, there has been a considerable interest in
characterizing and understanding the structural properties of
networked systems~\cite{AlBa02}. It has been established that
scale-free behavior~\cite{BaAl99} is one of the most fundamental
concepts for a  basic understanding of the organization of many
real-world systems in nature and society. This scale-free property
has a profound effect on almost every aspect on dynamic processes
taking place on networks~\cite{DoGoMe08}, including
robustness~\cite{AlJeBa00},
percolation~\cite{CaNeStWa00,CoErAvHa01},
synchronization~\cite{ArDiKuMoZh08}, games~\cite{SzFa07}, epidemic
spreading~\cite{PaVe01}, to name just a few. For instance, for a
wide range of scale-free networks an epidemic threshold does not
exist, and even infections with a low spreading rate will prevail
over the entire population in these networks~\cite{PaVe01}. This is
a radical change from the conclusions drawn from classical disease
modeling~\cite{He00}.

In addition to the above-mentioned dynamics, some authors have
focused their attention on the trapping problem occurring on complex
networks~\cite{Bobe05,CoBeTeVoKl07,KiCaHaAr08,CaAb08,ZhZhZhYiGu09,ZhGuXiQiZh09,ZhQiZhXiGu09},
which is one of the main topic of interest for random walks
(diffusion)~\cite{HaBe87,MeKl04}. The classical trapping problem
first introduced in~\cite{Mo69} is a random-walk issue, where a trap
is located at a fixed position, absorbing all particles that visit
it. An interesting quantity closely related to the trapping problem
is the mean first-passage time (MFPT), which is very important in
the study of transport-limited reactions~\cite{YuLi02,LoBeMoVo08},
and target search~\cite{BeCoMoSuVo05,Sh06}, amongst other physical
problems. A result from previous research is that a power-law
property can improve the efficiency of transport by diffusion on
scale-free
networks~\cite{Bobe05,ZhZhZhYiGu09,ZhGuXiQiZh09,ZhQiZhXiGu09}: the
MFPT, $\langle T \rangle$, scales linearly or sublinearly with the
number of network nodes $V$ as $\langle T \rangle \sim V^{\theta}$
with $\theta=1$ or $\theta<1$, which shows that the efficiency of
trapping processes on scale-free networks is even better than (at
least not worse than) that on complete graphs~\cite{Bobe05}, the
best possible structure for a fast diffusion (with $\langle T
\rangle \sim V$).

Although the scale-free topology has a direct effect on other
structural characteristics (e.g., average path length~\cite{CoHa03})
of networks and dynamics running on them, it cannot reflect all the
information of the network structure. Recently, it has been
discovered that many real-life networks, such as the WWW, metabolic
networks, and yeast protein interaction networks have self-similar
properties and exhibit fractal
scaling~\cite{SoHaMa05,SoHaMa06,GoSaKaKi06}. This fractal topology
is often characterized through the fractal dimension $d_B$, which
can be obtained by the box-counting
algorithm~\cite{SoGaHaMa07,GoHuDi08}. It is now commonly accepted
that fractal scaling~\cite{SoHaMa05} must be considered in an
integral basic understanding of the organization of real-life
complex systems.

As a fundamental property, topological fractality is related to many
respects of network structure and function. Recently, several
authors have shown that the correlation between degree and
betweenness centrality of nodes is much weaker in fractal network
models in comparison with non-fractal
models~\cite{KiHaPaRiPaSt07,ZhZhChGu08}. It has been also
shown~\cite{SoHaMa06,ZhZhZo07} that fractal scale-free networks are
disassortative~\cite{Newman02}, and this  feature, together with
fractality, makes such scale-free networks more robust against
intentional attacks on hub nodes, as compared to the highly
vulnerable non-fractal scale-free networks~\cite{SoHaMa06}. In
addition to the distinction in the robustness, fractal networks
exhibit lower synchronizability than their non-fractal
counterparts~\cite{ZhZhZo07}. Although a lot of efforts have been
devoted to fractal scale-free
networks~\cite{HiBe06,RoHaAv07,RoAv07,Hi07,ZhZhZoCh08}, it is still
of current interest to look for a better understanding of the
consequences of a fractal topology on different dynamic processes.

In this paper, we study the trapping dynamics on a fractal
scale-free network in the presence of a perfect absorber located at
a fixed node. We obtain a rigorous solution for the MFPT of the
unbiased random walks, which is computed through the recurrence
relations derived from the network structure. The resulting formula
shows that for a large network, the MFPT, $\langle T \rangle$,
scales with the network order $V$ as $\langle T \rangle \sim
V^{\frac{3}{2}}$. This superlinear growth is significantly different
from the linear or sublinear scaling previously found for nonfractal
scale-free networks.

\section{\label{sec:tree} The fractal scale-free network}

In this section we introduce a network model defined in an iterative
way~\cite{SoHaMa06}, which has attracted a great amount of
attention~\cite{CoBeTeVoKl07,RoHaAv07,ZhZhChGu08,GaSoHaMa07}. We
call this model {\em iterative fractal scale-free tree} (IFSFT). We
study  the IFSFT because of its intrinsic interest and its relevance
to real-world systems. For example, the so-called border tree motifs
have been shown to be present, in a significant way, in real-life
networks~\cite{ViroTrCo08}. Moreover, the IFSFT is deterministic,
which allows us to study analytically its topological properties and
dynamical processes taking place on it. It is thus a good test-bed
and an ideal substrate network.

\begin{figure}[h]
\begin{center}
\includegraphics[width=.45\linewidth,trim=120 110 100 120]{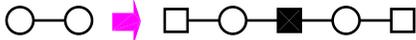}
\caption{(Color online) Iterative construction method of the
network. The next generation is obtained by performing the following
operation: for each edge, we replace it by a cluster on the
right-hand side of the arrow, where each $\square$ stands for a new
external node, while $\blacksquare$ represents an internal
node.}\label{construct}
\end{center}
\end{figure}

\begin{figure}
\begin{center}
\includegraphics[width=.6\linewidth,trim=80 45 70 0]{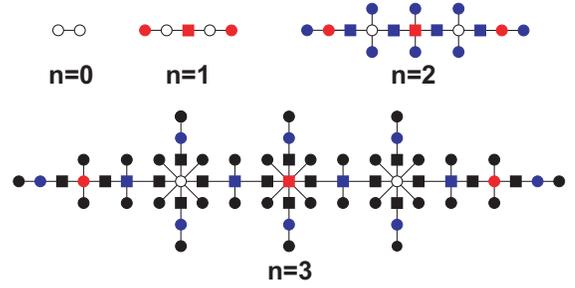} \
\end{center}
\caption[kurzform]{(Color online) The first three iterations of the
network.} \label{network}
\end{figure}

The IFSFT, denoted by $\mathbb{F}_{n}$ after $n$ ($n\geq 0$)
iterations (the number of iterations is also called generation
hereafter), is constructed as follows. For $n=0$, $\mathbb{F}_{0}$
is an edge connecting two nodes. For $n\geq 1$, $\mathbb{F}_{n}$ is
obtained from $\mathbb{F}_{n-1}$: for each edges in
$\mathbb{F}_{n-1}$, two new nodes (called external nodes with degree
of 1) are firstly introduced and linked respectively to both ends of
the edge; then, the edge is broken, another new node (referred to as
an internal node) is positioned in its middle and connected to both
ends (see Fig.~\ref{construct}). In other words, $\mathbb{F}_{n}$ is
obtained from $\mathbb{F}_{n-1}$ by performing the following
operations on every edge in $\mathbb{F}_{n-1}$: replace the edge by
a path of 2 links long, with the two endpoints of the path being the
same endpoints of the original edge, then attach a new node to each
endpoint of the path. Figure~\ref{network} shows the construction
process for the first three iterative processes.

According to the network construction, one can see that at each
generation $n_i$ ($n_i\geq 1$) the number of newly introduced nodes
is $L(n_i)=3\times4^{n_i-1}$. From this result, we can easily
compute the network order (i.e., the total number of nodes) $V_n$ at
generation $n$:
\begin{equation}\label{Nt}
 V_n=\sum_{n_i=0}^{n}L(n_i)=4^{n}+1.
\end{equation}

\begin{figure}
\begin{center}
\includegraphics[width=0.7\linewidth,trim=95 110 80 80]{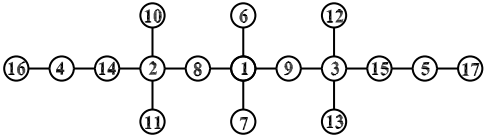}
\end{center}
\caption[kurzform]{\label{labeling} labels of all nodes of
$\mathbb{F}_2$.}
\end{figure}

To facilitate the description in what follows, we distinguish
different nodes of $\mathbb{F}_n$ by labeling them as follows. For
$\mathbb{F}_1$, the newly created internal node is labeled 1, the
initial two nodes belonging to $\mathbb{F}_0$ are labeled as 2 and
3, and the two new external nodes have labels 4 and 5, see
Fig.~\ref{labeling}. For each new iteration $n>1$, we label
consecutively the new nodes generated at this iteration, while we
keep the labels of the old nodes unchanged. Namely, new nodes are
labeled sequentially as $V_{n-1} + 1, V_{n-1} + 2, \ldots, V_n$. In
this way, we label each node by a unique integer: at generation $n$
all nodes are labeled from 1 to $V_n=4^{n}+1$. 

Let $k_i(n)$ be the degree of a node $i$ at generation $n$, which
entered the network at generation $n_i$ ($n_i\geq 0$). If node $i$
was an external node when it was introduced,
\begin{equation}\label{ki01}
k_i(n)=2^{n-n_{i}}\,;
\end{equation}
otherwise, if $i$ was an internal node when it entered the network,
\begin{equation}\label{ki02}
k_i(n)=2^{n-n_{i}+1}\,.
\end{equation}
From Eqs.~(\ref{ki01}) and~(\ref{ki02}), one can easily see that
after each new iteration the degree of a node doubles, i.e.,
\begin{equation}\label{ki03}
k_i(n)=2\,k_i(n-1)\,.
\end{equation}

The IFSFT presents some interesting topological characteristics. It
has a power law degree distribution with exponent
$\gamma=3$~\cite{RoHaAv07,ZhZhChGu08}. Its average path length,
defined as the mean of shortest distances between all pairs of
nodes, grows as a square power of the network
order~\cite{ZhZhChGu08}. The betweenness distribution of its nodes
exhibits a power-law behavior with exponent
$\gamma_b=\frac{3}{2}$~\cite{ZhZhChGu08}. Particulary, it is fractal
with the fractal dimension $d_B=2$.

After introducing the IFSFT, in what follows we will study the MFPT
for random walks with a single immobile trap on the network. We will
show that the fractal property has an essential influence on the
MFPT, i.e., the fractality can induce a general slowing down of
diffusion.

\section{Formulation of trapping problem on the IFSFT}

Here we formulate the trapping problem of a simple random walk of a
particle on the IFSFT $\mathbb{F}_n$ in the presence of an absorbing
trap positioned at the central hub node $1$, represented as $i_T$.
To this end, we first represent $\mathbb{F}_n$ by its adjacency
matrix $\textbf{A}_n$ of order $V_n\times V_n$. The entry $a_{ij}$
of $\textbf{A}_n$ is either 1 or 0: $a_{ij}=1$ if $i$ and $j$ are
adjacent and $a_{ij}=0$ otherwise. The diagonal degree matrix
$\textbf{D}_n$ of $\mathbb{F}_n$ is $\textbf{D}_n={\rm diag}
(k_1(n), k_2(n),\ldots, k_i(n), \ldots, k_{V_n}(n))$. Then, the
normalized Laplacian matrix of $\mathbb{F}_n$ is given by
$\textbf{L}_n=\textbf{I}_n-\textbf{D}_n^{-1}\,\textbf{A}_n$, where
$\textbf{I}_n$ is the $V_n\times V_n$ identity matrix.

In the trapping problem, at each time step, a particle, starting
from any node except the trap $i_T$, moves from its current location
to any of its nearest neighbors with equal probabilities. It is easy
to see that in the end the particle will be necessarily absorbed by
the trap, regardless of its starting location~\cite{Bobe05}. We are
interested in the mean transmit time (first-passage time, or
trapping time) $T_i^{(n)}$ for a particle, originating at node $i$,
to first reach the trap $i_T$ in $\mathbb{F}_n$.

Such a random walk can be described by a Markov chain~\cite{KeSn76},
whose fundamental matrix is the inverse of matrix $\mathbf{B}_n$
that is defined as a sub-matrix of the normalized Laplacian matrix
$\mathbf{L}_n$ obtained by deleting from it the first row and
column, corresponding to the absorbing node. The entry
$(b_n^{-1})_{ij}$ of the fundamental matrix $(\mathbf{B}_n)^{-1}$
expresses the mean residence time, which is the mean number of
visitations of node $j$ by the particle, starting from node $i$,
before trapping occurs. Thus, we have
\begin{equation}\label{MFPT4}
T_i^{(n)}=\sum_{j=2}^{V_n}(b_n^{-1})_{ij}\,.
\end{equation}
Then, the MFPT, $\langle T \rangle_n$, which is the mean of
$T_i^{(n)}$ over all nodes distributed uniformly over nodes in
$\mathbb{F}_n$ other than the trap, is given by
\begin{equation}\label{MFPT5}
 \langle T
\rangle_n=\frac{1}{V_n-1}\sum_{i=2}^{V_n}
T_i^{(n)}=\frac{1}{V_n-1}\sum_{i=2}^{V_n}\sum_{j=2}^{V_n}(b_n^{-1})_{ij}\,.
\end{equation}

\begin{figure}
\begin{center}
\includegraphics[width=0.35\linewidth,trim=115 20 125 45]{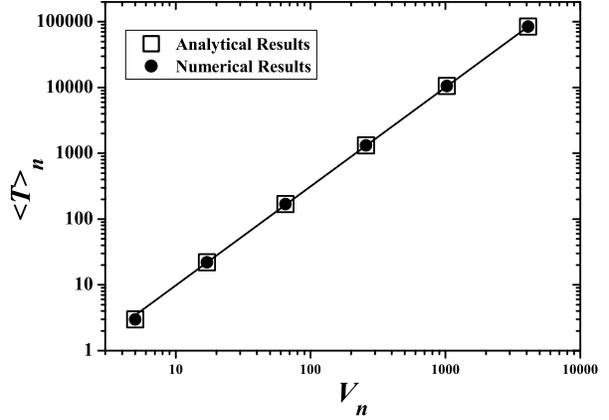}
\end{center}
\caption[kurzform]{\label{trap} Mean first-passage time $\langle T
\rangle_n$ versus network order $V_n$ on a log-log scale. The
numerical results are obtained by direct calculation from
Eq.~(\ref{MFPT5}), while the analytical results are from
Eq.~(\ref{MFPT19}). The solid line is the guide to the eye.}
\end{figure}

Equation~(\ref{MFPT5}) shows that the problem of finding $\langle T
\rangle_n$ is reduced to calculating the sum of all entries of the
fundamental matrix $(\mathbf{B}_n)^{-1}$. 
Although the expression of Eq.~(\ref{MFPT5}) seems compact, since
the order, $V_n-1$, of $(\mathbf{B}_n)^{-1}$ increases exponentially
with $n$, for large $n$, it becomes impossible to get $\langle T
\rangle_n$ through direct calculation from Eq.~(\ref{MFPT5}) as we
are restricted by time and computer memory, and one can calculate
directly the MFPT only for the first iterations, see
Fig.~\ref{trap}. However, the particular construction of the IFSFT
and the special choice of the trap location allow to calculate
analytically MFPT to obtain a closed-form formula. The derivation
details of which will be provided in the following section.

\section{Explicit expression for MFPT}

Before giving a general formula for MFPT, $\langle T \rangle_n$,
we first establish the dependence of $T_i^{(n)}$ on $n$. 

\subsection{Evolution scaling for trapping time}

For each $n$, the values of $T_i^{(n)}$ can be obtained
straightforwardly via Eq.~(\ref{MFPT4}). Table \ref{tab:AMTA2} lists
the numerical values of $T_i^{(n)}$ for nodes for the first several
generations up to $n=6$. The numerical values quoted in
table~\ref{tab:AMTA2} show that for a given node $i$ we have
$T_i^{(n+1)}=8\,T_i^{(n)}$. That is to say, upon growth of the IFSFT
from generation $n$ to $n+1$, the mean time to first reach the trap
increases by a factor of 8. This is a basic characteristic of random
walks on the IFSFT, which can be established from the arguments
below.

Consider an arbitrary node $i$ in the IFSFT $\mathbb{F}_n$. From
Eq.~(\ref{ki03}), we know that at iteration $n+1$, the degree of
node $i$ of IFSFT doubles, i.e., it grows from $k_i$ (degree at
iteration $n$) to $2\,k_i$. Moreover, all these $2\,k_i$ neighbors
are new nodes created at iteration $n+1$, among which $k_i$
neighbors are external nodes, and the rest $k_i$ neighbors are
internal nodes. We now examine the standard random walk in
$\mathbb{F}_{n+1}$: Let $A$ be the mean transmit time for a particle
starting from node $i$ to any of its $k_i$ old neighbors, i.e.,
those nodes directly linked to node $i$ at iteration $n$; and let
$B$ (resp. $C$) be the mean transmit time for going from any of the
$k_i$ internal (resp. external) neighbors of $i$ to one of the $k_i$
old neighbors. Then the mean transmit times follow the relations:
\begin{eqnarray}\label{MFPT6}
\left\{
\begin{array}{ccc}
A&=&\frac{1}{2}(1+C) + \frac{1}{2}(1+B)\,,\\
B&=&\frac{1}{2} + \frac{1}{2}(1+A)\,,\\
C&=&1+A\,.
 \end{array}
 \right.
\end{eqnarray}
Equation~(\ref{MFPT6}) has a solution $A=8$, found by eliminating
$B$ and $C$. That is to say, when the network grows from iteration
$n$ to iteration $n+1$, the first-passage time from any node $i$ to
any node $j$ (both $i$ and $j$ belong to $\mathbb{F}_n$) increases
by a factor of 8. Thus, we have $T_i^{(n+1)}=8\,T_i^{(n)}$, which
will be useful for the derivation of the exact formula for the MFPT
below.

\begin{table*}
\caption{The trapping time $T_i^{(n)}$ for a random walker starting
from node $i$ on the IFSFT for various $n$. Notice that owing to the
obvious symmetry, nodes in a parenthesis are equivalent, since they
have the same trapping time. All the values are calculated
straightforwardly from Eq.~(\ref{MFPT4}).} \label{tab:AMTA2}
\begin{center}
\begin{tabular}{l|cccccccccc}
\hline \hline  $n \backslash i$  &(2,3)&
(4,5)&(6,7)&(8,9)& (10,11,12,13) & (14,15) &(16,17)\\
\hline
            \hline
            1 & $3$ & $4$       \\
            2 & $24$ & $32$ & $1$ & $13$ & $25$         & $29$          & $33$ \\
            3 & $192$ & $256$ & $8$ & $104$ & $200$     & $232$         & $264$   \\
            4 & $1536$ & $2048$ & $64$ & $832$ & $1600$ & $1856$        & $2112$  \\
            5 & $12288$ & $16384$ & $512$ & $6656$ & $12800$ & $14848$ & $16896$  \\
            6 & $98304$ & $131072$ & $4096$ & $53248$ & $102400$ & $118784$ & $135168$  \\
\hline \hline
\end{tabular}
\end{center}
\end{table*}

\subsection{Formula for the MFPT}

After obtaining the scaling of mean trapping time for old nodes, we
now derive the analytical rigorous expression for the MFPT. Before
proceeding further, we first introduce some notation used in the
rest of this section. Let $\Delta_n$ denote the set of nodes in
$\mathbb{F}_n$, and let $\overline{\Delta}_n$ stand for the set of
those nodes introduced at generation $n$. To facilitate the
computation, we also define the following quantities for $m \leq n$:
\begin{equation}
    T_{m, {\rm tot}}^{(n)} = \sum_{i \in \Delta_m} T_i^{(n)},
\end{equation}
and
\begin{equation}
    \overline{T}_{m, {\rm tot}}^{(n)} = \sum_{i \in \overline{\Delta}_m} T_i^{(n)}.
\end{equation}

By definition, it follows that $\Delta_n = \overline{\Delta}_n \cup
\Delta_{n-1}$. Thus, we have
\begin{equation}\label{eq:Ttot}
T_{n, {\rm tot}}^{(n)} = T_{n - 1, {\rm tot}}^{(n)} +
\overline{T}_{n, {\rm tot}}^{(n)}
        = 8\,T_{n - 1, {\rm tot}}^{(n-1)} + \overline{T}_{n, {\rm
        tot}}^{(n)}\,,
\end{equation}
where the relation of $T_i^{(n+1)}=8\,T_i^{(n)}$ has been made use
of. Hence, to determine $T_{n, {\rm tot}}^{(n)}$, one should first
explicitly determine the quantity $\overline{T}_{n, {\rm
tot}}^{(n)}$. For this purpose, we further separate
$\overline{\Delta}_n$ into two sets: one set of external nodes and
the other set of internal nodes as defined in the second section,
which are denoted as $\overline{\Delta}_{n, {\rm ext}}$ and
$\overline{\Delta}_{n, {\rm int}}$, respectively. Clearly,
$\overline{\Delta}_n = \overline{\Delta}_{n, {\rm ext}} \cup
\overline{\Delta}_{n, {\rm int}}$. On the other hand, we have shown
that the cardinality of set $\overline{\Delta}_n$ is
$|\overline{\Delta}_n|=3\, \times 4^{n-1}$, and that $
|\overline{\Delta}_{n, {\rm ext}}| = 2|\overline{\Delta}_{n, {\rm
int}}|$. Thus, we can obtain $|\overline{\Delta}_{n, {\rm int}}| =
4^{n-1}$ and $|\overline{\Delta}_{n, {\rm ext}}| = 2 \times
4^{n-1}$. Then, two corresponding quantities can be defined:
\begin{equation}
    \overline{T}_{m, {\rm int}}^{(n)} = \sum_{i \in \overline{\Delta}_{m, {\rm int}}}
    T_i^{(n)}\,,
\end{equation}
\begin{equation}
    \overline{T}_{m, {\rm ext}}^{(n)} = \sum_{i \in \overline{\Delta}_{m, {\rm ext}}}
    T_i^{(n)}\,.
\end{equation}
It is obvious that
\begin{equation}\label{eq:TtotR}
    \overline{T}_{n, {\rm tot}}^{(n)} = \overline{T}_{n, {\rm int}}^{(n)} + \overline{T}_{n, {\rm
    ext}}^{(n)}\,.
\end{equation}

\begin{figure}[h]
\begin{center}
\includegraphics[width=.45\linewidth,trim=150 110 120 120]{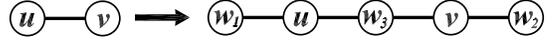}
\caption{Illustration showing the relation of the mean transmit
times for external and internal nodes.}\label{iter}
\end{center}
\end{figure}

In order to obtain $\overline{T}_{n, {\rm tot}}^{(n)}$, one may
alternatively get $\overline{T}_{m, {\rm int}}^{(n)}$ and
$\overline{T}_{m, {\rm ext}}^{(n)}$. To this end, we first establish
the relationship between the two quantities $\overline{T}_{m, {\rm
int}}^{(n)}$ and $\overline{T}_{m, {\rm ext}}^{(n)}$. By
construction, at a given generation, each edge connecting two nodes
$u$ and $v$ (see Fig.~\ref{iter}) will generate three new nodes in
the next generation: two external nodes ($w_1$ and $w_2$) and one
internal node ($w_3$), and the mean transmit times for these three
new nodes satisfy the following relations:
\begin{eqnarray}\label{MFPT61}
\left\{
\begin{array}{ccc}
T(w_1) &=&  1 +T(u)\,,\\
T(w_2) &=& 1+T(v)\,,\\
T(w_3) &=&
\frac{1}{2}\big(1+T(u)\big)+\frac{1}{2}\big(1+T(v)\big)\,.
 \end{array}
 \right.
\end{eqnarray}
Hence, we have
\begin{equation}\label{eq:intext1}
        T(w_1)+T(w_2) = 2\,T(w_3)\,.
\end{equation}
Summing Eq.~(\ref{eq:intext1}) over all old edges at the generation
before growth, we can easily obtain that for all $m\leq n$,
\begin{equation}\label{eq:intext2}
     \overline{T}_{m, {\rm ext}}^{(n)} = 2\,\overline{T}_{m, {\rm
     int}}^{(n)}\,.
\end{equation}
Equation (\ref{eq:intext2}) gives the relationship between
$\overline{T}_{n, {\rm ext}}^{(n)}$ and $\overline{T}_{n, {\rm
ext}}^{(n)}$, which is very significant since it is useful for the
computation in the following text.

Therefore, the issue of determining $\overline{T}_{n, {\rm
tot}}^{(n)}$ is reduced to finding the quantity $\overline{T}_{n,
{\rm ext}}^{(n)}$ that can be obtained as follows. For an arbitrary
external node $i_{\rm ext}$ in $\mathbb{F}_n$, which is created at
generation $n$ and attached to an old node $i$, we have
\begin{equation}\label{eq:intextX2}
   T_{i_{\rm ext}}^{(n)} = 1 + T_i^{(n)} \,,
\end{equation}
since a particle starting from node $i_{\rm ext}$ will be on node
$i$ after one jump. Note that Eq.~(\ref{eq:intextX2}) holds for any
node pair consisting of an old node and any one of its new external
adjacent nodes. By applying Eq.~(\ref{eq:intextX2}) to two sum (the
first one is over a given old node and all its new external nodes,
the other is summing the first one over all old nodes), we get
\begin{eqnarray}
    \overline{T}_{n, {\rm ext}}^{(n)} &=&|\overline{\Delta}_{n, {\rm ext}}| +
    \sum_{i \in \Delta_{n-1}}\left(k_i(n-1)\times T_i^{(n)}\right) \nonumber\\&=& |\overline{\Delta}_{n, {\rm ext}}| +
    \left(\overline{T}_{n-1, {\rm ext}}^{(n)} + 2\,\overline{T}_{n-1, {\rm int}}^{(n)}\right) \nonumber\\
    &\quad&  + \left(2\,\overline{T}_{n-2, {\rm ext}}^{(n)} + 4\,\overline{T}_{n-2, {\rm int}}^{(n)}\right) + \cdots  \nonumber\\
    &\quad& + \left(2^{n-2}\,\overline{T}_{1, {\rm ext}}^{(n)} + 2^{n-1}\, \overline{T}_{1, {\rm int}}^{(n)}\right) \nonumber\\
    &=& 2 \times 4^{n-1} + 2\,\overline{T}_{n-1, {\rm ext}}^{(n)} + 4\,\overline{T}_{n-2, {\rm ext}}^{(n)} + \cdots   \nonumber\\
    &\quad& + 2^{n-1}\,\overline{T}_{1, {\rm ext}}^{(n)}\,,  \label{eq:rec1}
\end{eqnarray}
where Eqs.~(\ref{eq:TtotR}) and~(\ref{eq:intext2}) were used.


Now, we can determine $\overline{T}_{n, {\rm ext}}^{(n)}$ through a
recurrence relation, which can be obtained easily. From
Eq.~(\ref{eq:rec1}), it is not difficult to write out
$\overline{T}_{n+1, {\rm ext}}^{(n+1)}$ as
\begin{equation}
    \overline{T}_{n+1, {\rm ext}}^{(n+1)} =
    2 \times 4^{n} + 2\,\overline{T}_{n, {\rm ext}}^{(n+1)} + 4\,\overline{T}_{n-1, {\rm ext}}^{(n+1)} + \cdots
     + 2^{n}\,\overline{T}_{1, {\rm ext}}^{(n+1)}\,.  \label{eq:rec2}
\end{equation}

Equation~(\ref{eq:rec2}) minus Eq.~(\ref{eq:rec1}) times 16 and
making use of the relation $T_i^{(n+1)}=8\,T_i^{(n)}$, one gets the
following recurrence relation
\begin{equation} \label{eq:T1}
    \overline{T}_{n+1, {\rm ext}}^{(n+1)} = 32\,\overline{T}_{n, {\rm ext}}^{(n)} - 6 \times 4^n\,.
\end{equation}
Considering the initial condition $\overline{T}_{2, {\rm ext}}^{(2)}
= 168$, this recurrence relation is solved to obtain
\begin{equation} \label{eq:TX1}
\overline{T}_{n, {\rm ext}}^{(n)} = \frac{3}{56}(4^{n + 1} + 3
\times 32^n)\,.
\end{equation}
Inserting Eq.~(\ref{eq:TX1}) into Eq.~(\ref{eq:TtotR}) and
considering the relation $\overline{T}_{n, {\rm ext}}^{(n)}=2\,
\overline{T}_{n, {\rm int}}^{(n)}$, we have
\begin{equation}\label{eq:Tnew}
    \overline{T}_{n, {\rm tot}}^{(n)} =\frac{3}{2}\,\overline{T}_{n, {\rm ext}}^{(n)}= \frac{9}{112}(4^{n + 1} + 3 \times 32^n)\,.
\end{equation}
Substituting the last expression into Eq.~(\ref{eq:Ttot}) yields
\begin{equation}
T_{n, {\rm tot}}^{(n)} = 8\,T_{n-1, {\rm tot}}^{(n-1)} +
\frac{9}{112}(4^{n + 1} + 3 \times 32^n)\,.  \label{eq:T2}
\end{equation}
Using $T_{1, {\rm tot}}^{(1)} = 14$, Eq.~(\ref{eq:T2}) is solved
inductively
\begin{equation}\label{eq:Tnew1}
T_{n, {\rm tot}}^{(n)} = \frac{9}{28}(32^n - 4^n) + \frac{35}{56}
\times 8^n\,.
\end{equation}

Inserting Eq.~(\ref{eq:Tnew1}) into Eq.~(\ref{MFPT5}), we obtain the
rigorous expression for the MFPT for the trapping problem on the
$n$-th generation of the IFSFT:
\begin{equation}\label{MFPT19}
 \langle T
\rangle_n =\frac{9}{28}(8^n -1) + \frac{35}{56} \times 2^n\,.
\end{equation}
We have checked this exact solution for the MFPT against numerical
values given by Eq.~(\ref{MFPT5}), see Fig.~\ref{trap}. For all $1
\leq n \leq 6$, the analytical values obtained from
Eq.~(\ref{MFPT19}) are perfectly consistent with the numerical
results. This agreement is an independent test of our theoretical
formula.

We show next how to represent MFPT as a function of the network
order, with the aim of obtaining the relation between these two
quantities. Recalling Eq.~(\ref{Nt}), we have $4^{n}=V_n-1$ and
$n=\log_4\big(V_n-1\big)$. Hence, Eq.~(\ref{MFPT19}) can be recast
as
\begin{equation}\label{MFPT20}
\langle T \rangle_n = \frac{9}{28}\left((V_n-1)^{\frac{3}{2}}
-1\right) + \frac{35}{56} \times (V_n-1)^{\frac{1}{2}}\,.
\end{equation}
For a large network, i.e., $V_n\rightarrow \infty$,
\begin{equation}\label{MFPT21}
\langle T \rangle_n \sim (V_n)^{\frac{3}{2}}\,,
\end{equation}
with the exponent $\frac{3}{2} =1.5$ much larger than 1. Thus, in
the limit of large network order $V_n$, the MFPT grows superlinearly
with the number of network nodes.

Recently, it has been shown that for non-fractal scale-free networks
with a large network order $V$, their MFPT $\langle T \rangle$
behaves linearly or sublinearly with $V$ as $\langle T \rangle \sim
V^{\theta}$ with $\theta \leq
1$~\cite{Bobe05,ZhZhZhYiGu09,ZhGuXiQiZh09,ZhQiZhXiGu09}. However, we
have seen that for the IFSFN, the MFPT, $\langle T \rangle_n$,
increases superlinearly with $V_n$ (i.e., $\langle T \rangle_n \sim
(V_n)^{\frac{3}{2}}$) irrespective of its scale-free property,
presenting an obvious difference from the results previously
obtained for its non-fractal scale-free counterparts.

Why is the MFPT for the IFSFT far larger than that for nonfractal
scale-free networks such as the Apollonian
network~\cite{ZhGuXiQiZh09}? The reasons behind this discrepancy may
be explained as follows. For the Apollonian network, the
large-degree nodes, including the trap node, are directly connected
to one another and compose a core group sharing more neighbors,
which make the Apollonian network be a very compact system. So,
these large-degree nodes can be easily visited by a particle in
spite of its starting location. The interconnection within nodes
with large degrees makes the particle spend a short time to find the
trap. On the contrary, in the IFSFT, the large-degree nodes are not
linked to each other, they are exclusively connected to low-degree
nodes, as a result from the fractal property of IFSFT. In other
words, there is an effective `repulsion' between the large-degree
nodes in the IFSFT~\cite{SoHaMa06,ZhZhChGu08}, which seems to be a
main feature that distinguishes the IFSFT from the Apollonian
network. Hence, for the trapping problem in the IFSFT, because of
the isolation of the large-degree nodes from each other, the
particle will first reach a hub node, and before being absorbed it
will spend a lot of time in the intermediate regions constituted by
small-degree nodes, which connect indirectly the large-degree nodes
to one another. Therefore, it takes a longer time for the particle
to arrive at the trap.

\section{Conclusions}

We have studied a classical trapping problem performed on a
deterministically growing scale-free network with fractal topology.
The self-similarity of the network allows us to derive the exact
expression for the MFPT. We have shown that, for large networks, the
MFPT increases as a power-law function of the network order, with an
exponent much larger than 1, in contrast with previous results found
for non-fractal scale-free networks. We see that this slow transport
efficiency lies with the inherent fractality and its associated
disassortativity. Thus, we can infer that scale-free networks do not
always tend to accelerate the diffusion processes occurring on them.
Our research may be helpful for a better understanding of the role
that network structure plays in a trapping process.

\section{Acknowledgment}

We thank F. Comellas for his help in revising the manuscript. This
research was supported by the National Basic Research Program of
China under grant No. 2007CB310806, the National Natural Science
Foundation of China under Grant Nos. 60704044, 60873040 and
60873070, Shanghai Leading Academic Discipline Project No. B114, and
the Program for New Century Excellent Talents in University of China
(NCET-06-0376). W L Xie also acknowledges the support provided by
Hui-Chun Chin and Tsung-Dao Lee Chinese Undergraduate Research
Endowment (CURE).

\end{document}